\documentclass[a4paper,twocolumn,prb,showkeys]{revtex4}
\usepackage{amsmath}    % need for subequations
\usepackage{graphicx}   % need for figures
\usepackage{verbatim}   % useful for program listings
\usepackage{color}      % use if color is used in text
\usepackage{subfigure}  % use for side-by-side figures
\usepackage{hyperref}   % use for hypertext links, including those to external documents and URLs

\begin{document}

\title{Anomalous Magnetoresistance in Fibonacci Multilayers}

\author{L.D. Machado$^1$, C.G. Bezerra$^1$, M.A. Correa$^1$, C. Chesman$^{1,2}$, J.E. Pearson$^2$ and A. Hoffmann$^2$}

\address{$^1$Departamento de F\'\i sica Te\'orica e Experimental, Universidade Federal do Rio  Grande do Norte, Natal-RN 59072-970, Brazil}

\address{$^2$Materials Science Division, Argonne National Laboratory, Argonne, IL, 60439, USA}

\date{\today}

\begin{abstract}
  The present paper theoretically investigates magnetoresistance curves in quasiperiodic magnetic multilayers for two different growth directions, namely [110] and [100]. We considered identical ferromagnetic layers separated by non-magnetic layers with two different thicknesses chosen based on the Fibonacci sequence. Using parameters for Fe/Cr multilayers, four terms were included in our description of the magnetic energy: Zeeman, cubic anisotropy, bilinear and biquadratic couplings. The minimum energy was determined by the gradient method and the equilibrium magnetization directions found were used to calculate magnetoresistance curves. By choosing spacers with a thickness such that biquadratic coupling is stronger than bilinear coupling, unusual behaviors for the magnetoresistance were observed: (i) for the [110] case there is a different behavior for structures based on even and odd Fibonacci generations; and more interesting, (ii) for the [100] case we found magnetic field ranges for which the magnetoresistance increases with magnetic field.
\end{abstract}


\maketitle

\section{Introduction}

The study of magnetic multilayers has been the focus of much attention since the discovery of antiferromagnetic bilinear coupling between magnetic Fe layers separated by nonmagnetic Cr layers\cite{Grunberg}. The subsequent discovery of giant magnetoresistance (GMR)\cite{Fert}, which allowed the electrical resistance in these systems to be controlled through external magnetic fields, led to several applications, particularly in the field of information storage. In 1990 Parkin \textit{et al.} \cite{Parkin} showed that, depending on spacer thickness, bilinear coupling between magnetic films oscillated between being ferromagnetic and antiferromagnetic. One year later, R\"uhrig \textit{et al.} \cite{Ruhrig} discovered a novel form of coupling (later called biquadratic coupling) in which, for certain spacer thickness, non-collinear coupling existed between the magnetic films, resulting in a $90^{\circ}$ angle between magnetization of adjacent films.

Around the same time, other important breakthroughs were being made in what was then an unrelated field. The discovery of quasicrystals by Shechtman \textit{et al.} \cite{Shechtman} in 1984 confirmed the existence of an intermediate phase between ordered crystals and disordered solids. A year later, Merlin \textit{et al.} \cite{Merlin} reported performing the first quasiperiodic superlattice following the Fibonacci sequence. More recently, first quasiperiodic Fe/Cr magnetic multilayers with biquadratic coupling were grown experimentally \cite{Thatiara}, illustrating the development of crystal growth techniques, which allow substantial thickness control for each layer.

The magnetic properties of multilayers can depend significantly on the stacking pattern of their layers, which can now be tailored in unusual stacking arrangements. For instance, a quasiperiodic stacking pattern in Fe/Cr magnetic multilayers induces new magnetic phases which would not be observed in a periodic arrangement. The consequences of these new phases are observed in the static \cite{PRB99} and dynamic properties \cite{JPCM2002} of these magnetic structures. As previously mentioned, the spacer thickness greatly influences the property of these multilayers; a relevant question that naturally arises is: what are the consequences of a quasiperiodic stacking pattern of the non-magnetic spacers? This paper investigates a new stacking pattern with varying spacer thickness. In our model the spacer can have one out of three different thicknesses, which results in variations of the relative strength of bilinear and biquadratic couplings. The non-magnetic layers are arranged in a Fibonacci quasiperiodic sequence, and interesting properties emerge for specific combinations of spacers. Furthermore, results were obtained for two different growth directions - [100] and [110].

This paper is organized as follows. In Sec.\ II we discuss the theoretical model, with emphasis on the description of the quasiperiodic sequences and the crystallographic orientations considered here. The numerical method, used to obtain the equilibrium configurations, is described in Sec.\ II, as well. The numerical results are described in Sec.\ III  and our findings are summarized in Sec.\ IV.

\section{Theory}

A quasiperiodic multilayer can be built by juxtaposing two building blocks ($A,B$) following a quasiperiodic sequence. The Fibonacci sequence is widely used, with building blocks transforming according to the following rule: $A \rightarrow AB$, $B \rightarrow A$. The first Fibonacci sequence is $S_1=A$, the second is $S_2=AB$, the third is $S_3=ABA$ and so on. A more detailed description of quasiperiodic sequences can be found in the Ref.\ [10].

\begin{figure}
\centering
\includegraphics[scale=0.25]{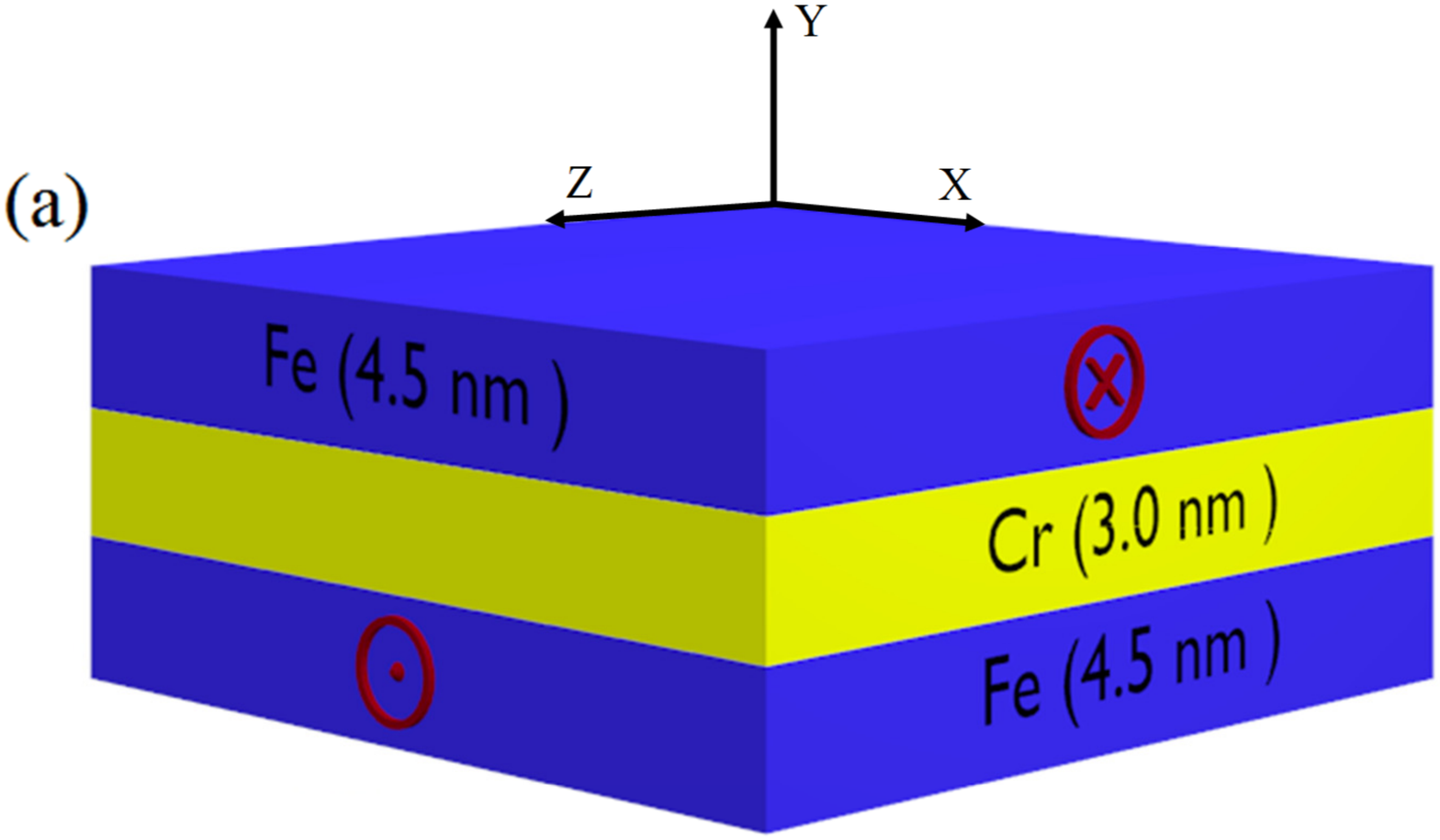}
\includegraphics[scale=0.25]{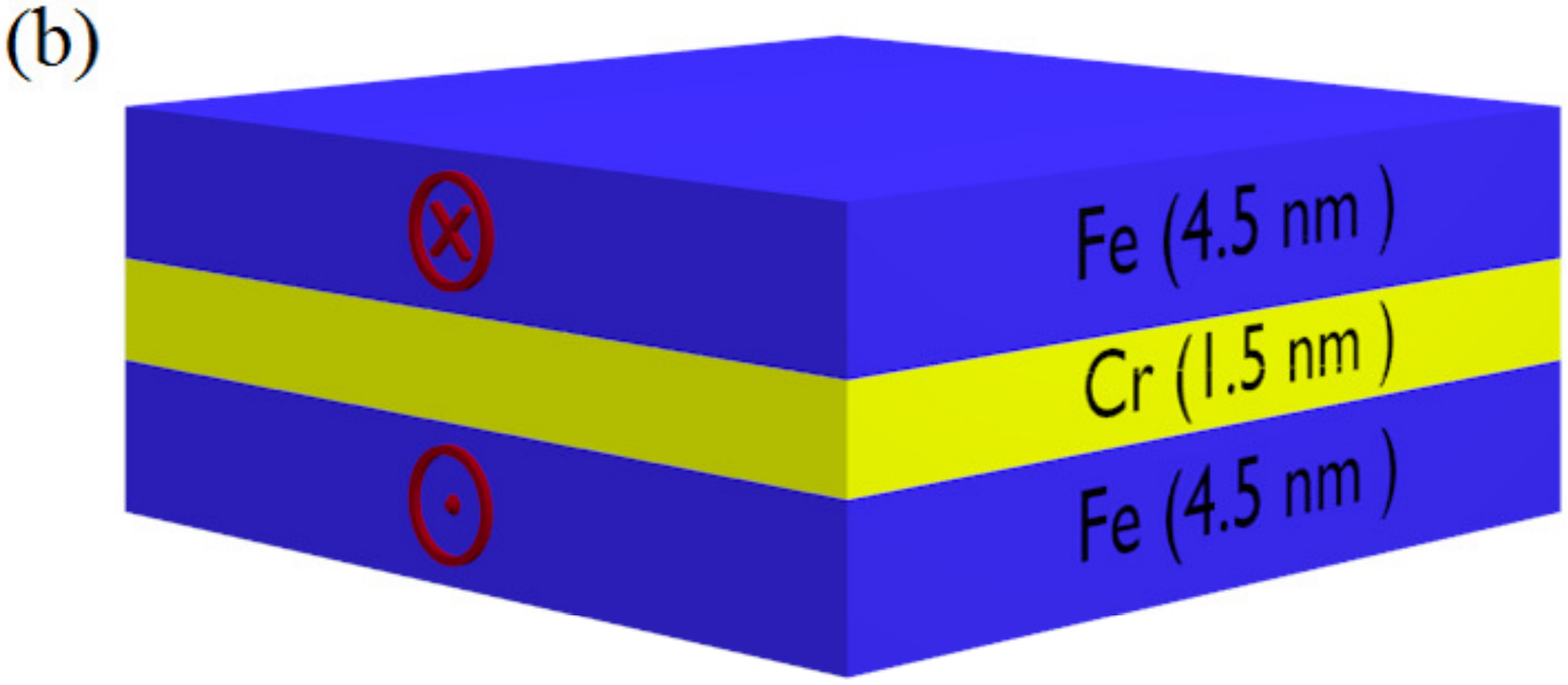}
\includegraphics[scale=0.25]{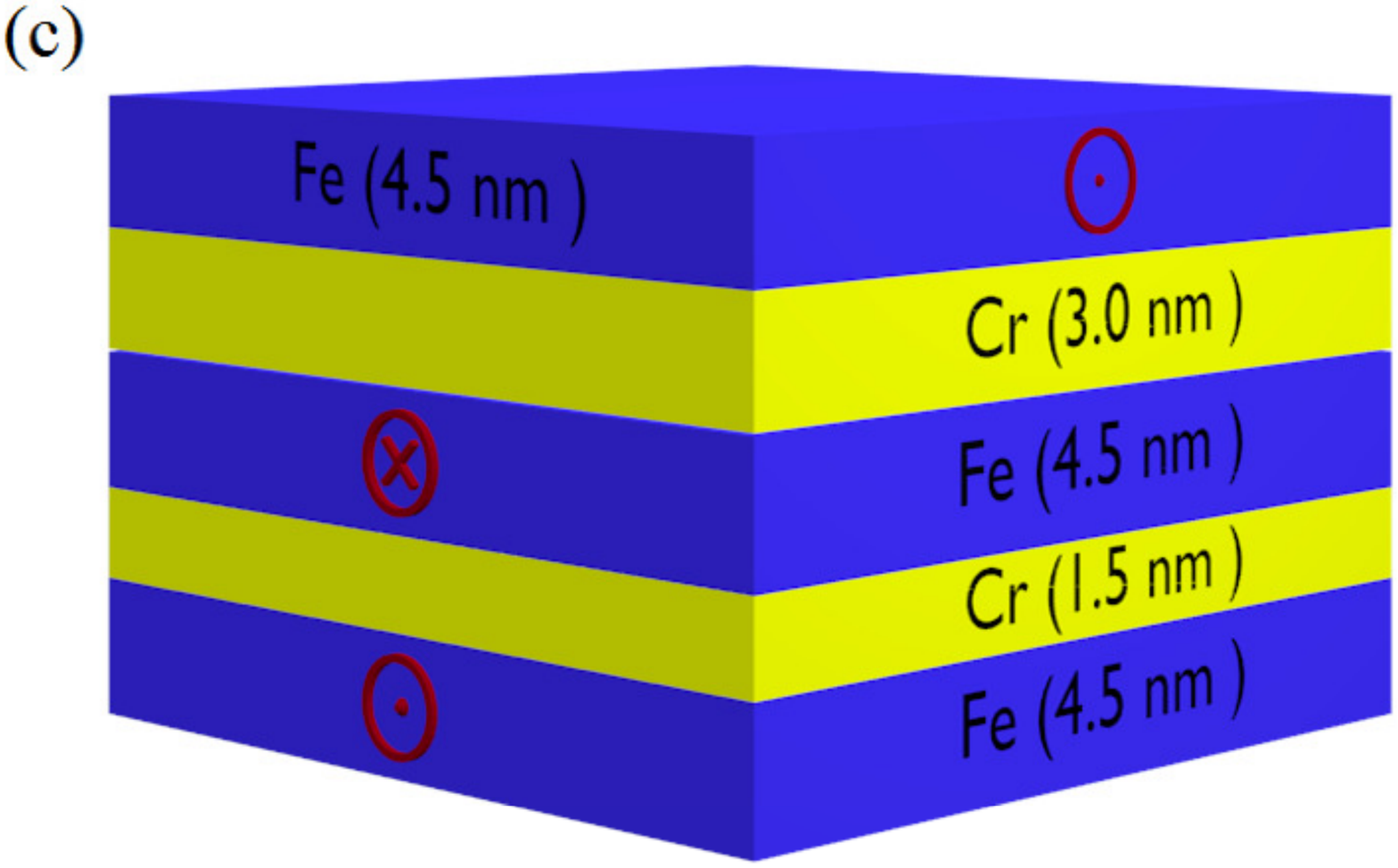}
\includegraphics[scale=0.25]{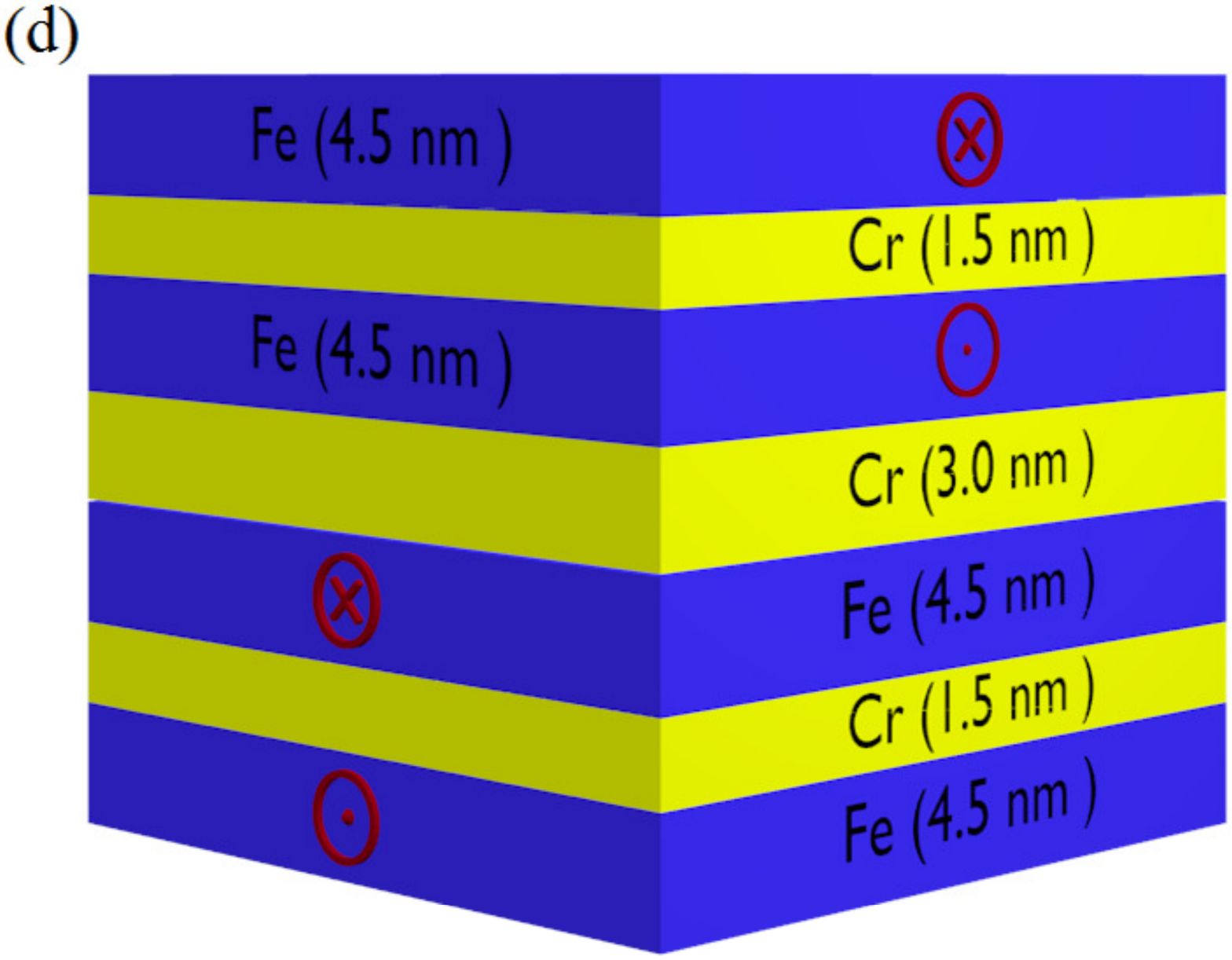}
\caption {Schematic multilayers constructed following the Fibonacci sequence. (a) and (b) correspond to $S_1$, one representing Cr thickness equal to $t_1=3.0$ nm and the other for Cr thickness equal to $t_2=1.5$ nm. (c) $S_2$ and (d) $S_3$ depict the magnetic counterpart of the second and third Fibonacci sequence, respectively.}
\label {fig1}
\end{figure}

In the present study, non-magnetic Cr layers, between ferromagnetic Fe layers, were chosen with thicknesses following the Fibonacci sequence. $A$ is a Cr layer with thickness $t_1$ and $B$ is a Cr layer with thickness $t_2$. For instance, the multilayer $Fe/Cr(t_1)/Fe/Cr(t_2)/Fe/Cr(t_1)/Fe$, corresponds to $Fe/A/Fe/B/Fe/A/Fe$. Illustrations of multilayers with non-magnetic layers following sequences $S_1$, $S_2$, $S_3$ and $S_4$, are shown in Fig.\ \ref{fig1}.

In order to describe the magnetic behavior of these multilayer systems, we considered four terms in the magnetic energy: the Zeeman term (owing to interaction between the magnetization of the ferromagnetic films and the applied external magnetic field), the cubic anisotropy term (due to interaction between the crystalline structure and electronic spins) and the two aforementioned terms that couple the magnetization of Fe layers separated by Cr layers, namely bilinear and biquadratic couplings. Considering these terms, the total magnetic energy can be written as \cite{Mariz},
$$\frac{E_T}{dM_S}=\displaystyle\sum_{i=1}^{n}\left[-H_0\cos(\theta_i-\theta_H)+\frac{H_{ac}}{8}\sin^2(2\theta_i) \right]$$
\begin{equation}
+\displaystyle\sum_{i=1}^{n-1}\left[-H_{bl_i}\cos(\theta_i-\theta_{i+1})+H_{bq_i}\cos^2(\theta_i-\theta_{i+1}) \right]
\label{Et010_EV}
\end{equation}
for the [100] direction, and
$$\frac{E_T}{dM_S}=\displaystyle\sum_{i=1}^{n}\left[-H_0\cos(\theta_i-\theta_H)+\frac{H_{ac}}{8}\left(\cos^4\theta_i+  \sin^2 2\theta_i \right) \right]$$
\begin{equation}
+\displaystyle\sum_{i=1}^{n-1}\left[-H_{bl_i}\cos(\theta_i-\theta_{i+1})+H_{bq_i}\cos^2(\theta_i-\theta_{i+1}) \right]
\label{Et110_EV}
\end{equation}
for the [110] direction. A comparison of the two equations shows that the cubic anisotropy terms depends on the growth direction. A thorough description of how this term is calculated for both growth directions can be found in Ref.\ [11]. In these equations $d$ represents the thickness of the Fe layers (which in our model is constant), $M_S$ is the saturation magnetization, $n$ is the total number of ferromagnetic films, $H_0$ is the external magnetic field that we consider to be maintained within the plane of the films (in our case the x-z plane, see Fig.\ \ref{fig1}), $\theta_H$ is the angle between the external magnetic field and the z axis, $H_{bl}$ is the bilinear coupling term that gives rise to parallel (anti-parallel) magnetization alignment of adjacent Fe films if positive (negative), and $H_{bq}$ is the biquadratic coupling term aligning the magnetization of adjacent Fe films in a perpendicular manner. $H_{ca}$ measures the strength of the cubic anisotropy field and, for the [100] case, tends to align magnetization of the films parallel to the crystalline axis (either x or z), whereas in the [110] case the magnetization tends to be aligned parallel to the x direction (although there is a local minimum along the z direction, and a maximum at $\theta\approx35^{\circ}$). In accordance with the values given by Refs.\ [12,13], we used the numerical value of $H_{ca}= 0.5$ kOe and selected $\theta_H=0$ for both cases (this means the field is applied in the easy axis for the [100] case and the intermediate axis in the [110] case).

Another important aspect of these equations is that the bilinear and biquadratic fields change from one pair of layers to the next. This is due to the varying spacer thicknesses since, as previously mentioned, the values of these coupling terms strongly depend on this thickness. We performed calculations for three different values of spacer thickness:
\begin{enumerate}
\item $t=1.0$ nm \, for which $H_{bq}=0.1|H_{bl}|$ with $H_{bl}=-1.0$ kOe;
\item $t=1.5$ nm \, for which $H_{bq}=0.3|H_{bl}|$ with $H_{bl}=-0.15$ kOe;
\item $t=3.0$ nm \, for which $H_{bq}=|H_{bl}|$ with $H_{bl}=-0.035$ kOe.
\end{enumerate}
These values are the same as those found in Refs.\ [12,13]. In general, if we choose the second set for Cr layers that correspond to $A$ and the third set for Cr layers that correspond to $B$, we obtain different results from those we would have obtained if we had chosen the third set for $A$ and the second set for $B$. This means there is a total of six sets of parameters. We found more interesting results for the case where the biquadratic is relatively strong.

In order to calculate magnetoresistance for these multilayer systems, we need the set $\{ \theta_i \}$ of equilibrium angles that minimize equation \ref{Et010_EV} (or \ref{Et110_EV}). As the number of ferromagnetic films rises, the computational cost of numerically minimizing these equations increases, requiring an efficient method of calculating this minimum. As such, we applied the gradient method, which takes into account the gradient of $E_T$ in relation to the set $\{\theta_i \}$,
\begin{equation}
\vec{\nabla}E_T = \sum_{i=1}^n {\partial E_T\over
\partial\theta_i} \hat{\theta}_i.
\end{equation}

\noindent A brief description of this algorithm is:
\begin{enumerate}

\item[(i)] An initial set of angles was randomly chosen, $\{\theta_i\}_0$. These were used to calculate an initial energy $E_0$;

\item[(ii)] The gradient of the magnetic energy was calculated, $\vec{\nabla}E_T$, and the set $\{\theta_i\}_0$ was employed to find its numerical value
$\vec{\nabla}E_T(\{\theta_i\}_0)$;

\item[(iii)] We applied the calculated energy and gradient to find the next set of angles $\{\theta_i\}_1$, using $\{\theta_i\}_1=\{\theta_i\}_0-\alpha\vec{\nabla}_iE_T(\{\theta_i\}_0)$ for each ferromagnetic film;

\item[ (iv) ] The energy $E_1$ was then calculated based on this new set of angles. If $E_1<E_0$ this energy and the new set of angles were stored, otherwise we halved the value of $\alpha$ and repeated step (iii);

\item[ (v) ] This process was repeated until $\alpha$ was smaller than a given tolerance.

\end{enumerate}
\noindent A complete discussion of this method can be found in Ref.\ [10].

Theoretically, it is well known that spin-dependent scattering is responsible for the magnetoresistance ($M_R$) effect in these multilayers \cite{Gallagher}. It was also shown that $M_R$ varies linearly with $\cos(\Delta \theta)$ when electrons form a free-electron gas, i.e., there are no barriers between adjacent films \cite{Vedy}. Here, $\cos(\Delta \theta)$ is the angular difference between adjacent film magnetizations. In metallic systems such as Fe/Cr this angular dependence is valid and once the set $\{ \theta_i \}$ of equilibrium angles is determined, we
obtain normalized values for magnetoresistance \cite{PRB99}, i.e.,

\begin{equation}
M_R(H_0)=R(H_0)/R(0)=\frac{\displaystyle\sum_{i=1}^{n-1}\left[1-\cos(\theta_i-\theta_{i+1})\right]}{2(n-1)},
\label{mr}
\end{equation}
where $R(0)$ is the electric resistance at zero field.

\section{Numerical Results}

Although calculations were performed with several different sets of parameters, the remainder of this paper focuses on only one of these, since we determined this is sufficient to illustrate our system's most relevant properties. We selected the second set of parameters for Cr films associated with $A$ letters of the quasiperiodic sequence and the third set of parameters for Cr films associated with $B$ letters of the quasiperiodic sequence. From now on, we label them as Cr($A$) and Cr($B$), respectively.

\subsection{[110] cubic anisotropy}

Let us discuss our numerical results for the magnetoresistance in the case of the [110] growth direction. These results are illustrated in Figs.\ \ref{fig2} and \ref{fig3}. In Fig.\ 2 we present the magnetoresistance considering the Cr layers following the fourth and sixth Fibonacci generations, which means 6 and 14 Fe films, respectively. As we can see, all transitions are first-order type, characterized by discontinuous jumps in the magnetoresistance. For the fourth generation of the Fibonacci sequence ($S_4=ABAAB$), which is illustrated in Fig.\ 2(a), in the small field region, the magnetoresistance value is $1$ because all magnetizations are antiparallel to each other at zero field. As the external magnetic field increases ($\sim 80$ Oe), a transition takes place to a magnetic phase in which the magnetization of the bottom layer is aligned with the field. We can observe that, increasing the magnetic field, more transitions take place and the saturated phase emerges when $H_{0}\geq 570$ Oe. A similar behavior is observed for the sixth generation of the Fibonacci sequence ($S_6=ABAABABAABAAB$) which is shown in Fig.\ 2(b). As in the fourth generation case, in the low field region the magnetizations are in the antiferromagnetic configuration. As the field increases, ten different transitions are observed, from the antiferromagnetic configuration ($H_{0}<90$ Oe) to the saturated regime ($H_{0}\geq 570$ Oe). It is easy to note the self-similar pattern, which is the basic signature of a quasiperiodic system, present in Fig.\ \ref{fig2}, i.e., the magnetoresistance profile of the fourth generation is reproduced in the magnetoresistance profile of the sixth generation. Let us now take a look at the results for the magnetoresistance considering the Cr layers following the fifth ($S_5=ABAABABA$) and seventh ($S_7=ABAABABAABAABABAABABA$) Fibonacci generations, which means 9 and 22 Fe films, respectively. These results are illustrated in Fig.\ \ref{fig3}. Once again, there is a clear self-similar pattern which is shown Fig.\ \ref{fig3}, i.e., the magnetoresistance profile of the fifth generation is reproduced in the magnetoresistance profile of the seventh generation. One can observe that in the low field region the central magnetoresistance step is much larger than the case of even generations. This is because the even generations of the Fibonacci sequence are terminated by $B$. This letter is associated with the third set of parameters (lower values of $H_{bl}$ and $H_{bq}$). As a consequence, the Fe film at the bottom of the multilayer is weakly coupled to its only adjacent Fe film, and a lower magnetic field is enough to induce a transition. Therefore, if we compare Figs.\ \ref{fig2} and \ref{fig3}, we can see that structures built using even and odd Fibonacci generations present different profiles for the magnetoresistance. Moreover, we can also remark that there are two self-similar patterns: one for the even generations and another for the odd generations. As explained above, this is also a consequence of the subtle difference between even and odd Fibonacci generations. Such behavior had been observed previously in the specific heat of quasiperiodic magnetic superlattices \cite{bezerra1}.

\subsection{[100] cubic anisotropy}

Fig.\ \ref{fig4} depicts the (a) fourth and (b) fifth Fibonacci generations obtained for growth direction [100]. It illustrates a number of interesting properties. As in the [110] case, there are various first-order phase transitions, which are proportional to the number of ferromagnetic layers. Much more interesting, however, is the behavior of the magnetoresistance in the low magnetic field region. Fig.\ \ref{fig4} shows, for both fourth and fifth generations, a region where one can see a positive change of the magnetoresistance, i.e., a region where an increase in the magnetic field leads to a rise in magnetoresistance, that is, $\Delta M_R/\Delta H > 0$. In order to understand these positive changes in magnetoresistance, it is necessary to analyze the magnetization behavior of the various films. Fig.\ \ref{fig5} shows a diagram of the fourth Fibonacci generation, illustrating the magnetization direction of each ferromagnetic layer. The numbered arrows indicate the magnetization direction of different layers. For example, number 1 represents the first layer, on the top, and number 6 represents the last layer, on the bottom, of the multilayer. Once the cubic anisotropy is dominant, all magnetizations remain close to a crystalline axis, as observed in the diagram. For low magnetic field, the Zeeman term is not important and it can be ignored. As a consequence the film magnetizations tend to form a configuration that minimizes the bilinear and biquadratic energies. For Cr($A$), the sum of the two terms is minimized when  $\theta_i- \theta_{i+1} = 180^{\circ}$, whereas for Cr($B$), the sum of the two terms is minimized when  $\theta_i- \theta_{i+1} = 90^{\circ}$. In Fig.\ \ref{fig5} one can observe that in the low field region the film magnetizations are not in the anti-parallel configuration because of two Cr($B$) spacers in the multilayer. As the magnetic field increases, the Zeeman energy plays a more important role. A transition takes place for $H\sim 46$ Oe. For this configuration, all magnetizations, except for the bottom film magnetization, are in the anti-parallel configuration. Thus, the magnetoresistance increases, resulting in a transition with $\Delta M_R/\Delta H > 0$. When the magnetic field reaches $90$ Oe, a second transition takes place. In this configuration only the fourth Fe film changes its magnetization anti-parallel to the magnetic field. Therefore, the magnetoresistance drops to $\sim0.64$. With further increase of the magnetic field, it becomes energetically too costly for the film magnetizations to be opposite to the external field. This implies that the next transition, which takes place for $H\sim130$ Oe, leads to a configuration for which there is no film magnetization anti-parallel to the external magnetic field. However, most of magnetizations are orthogonal to each other. As a consequence, according to Eq.\ \ref{mr}, the magnetoresistance of this configuration is higher than the previous one. Once again, we observe a transition with $\Delta M_R/\Delta H > 0$. As the magnetic field increases, the film magnetizations gradually become aligned with the field, and the magnetoresistance monotonically decreases with the magnetic field. Saturation is reached for $H_S\sim450$ Oe. An analogous analysis applies to the fifth generation of the Fibonacci sequence depicted in Fig.\ \ref{fig4}b.

\section{Conclusion}

In summary, we studied quasiperiodic magnetic multilayers, composed by ferromagnetic Fe layers separated by non-magnetic Cr layers. The non-magnetic Cr layers were arranged according to the Fibonacci quasiperiodic sequence, such that the letters A and B in the sequence correspond to Cr layers with thicknesses $t_1$ and $t_2$, respectively. The Fe layers are between Cr layers as well as on the top and bottom of the multilayer structure. The calculation is based on a phenomenological model which includes the following contributions to the magnetic energy: Zeeman, cubic anisotropy, bilinear and biquadratic exchanges. The magnetic energy was minimized using the gradient method and the resulting equilibrium angles were used to calculate magnetoresistance curves for the system. We selected a particular set of parameters such that the thickness of Cr($A$) layer corresponds to $H_{bq} = 0.3 |H_{bl}|$ and the thickness of Cr($B$) layer corresponds to $H_{bq} = |H_{bl}|$. These two sets of exchange couplings are responsible for the exchange energies between two adjacent Fe films. We numerically calculated the magnetoresistance curves assuming two possible crystallographic orientations namely, [110] and [100]. Our results show that quasiperiodic magnetic multilayers exhibit a rich variety of configurations induced by the external magnetic field. In particular, two points may be emphasized: (i) the well-defined even and odd parity observed in the behavior of the magnetoresistance curves and (ii) the positive change of the magnetoresistance with $\Delta M_R/\Delta H > 0$.

\begin{figure}
\centering
\includegraphics[scale=1.5]{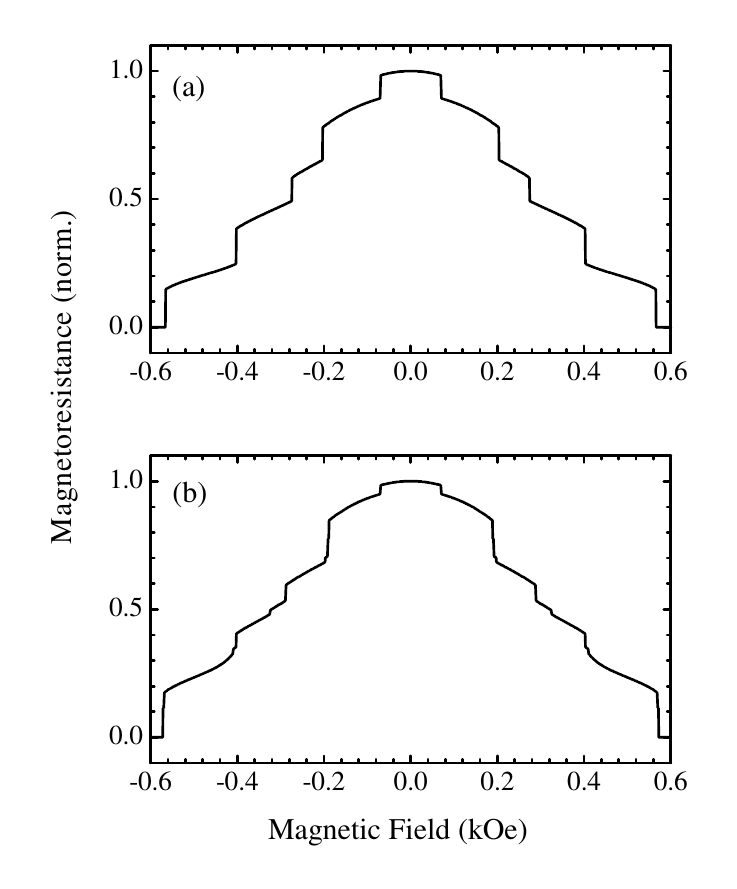}
\caption {Normalized magnetoresistance curves for growth direction [110] for the (a) fourth and (b) sixth Fibonacci generations.}
\label {fig2}
\end{figure}

\begin{figure}
\centering
\includegraphics[scale=1.5]{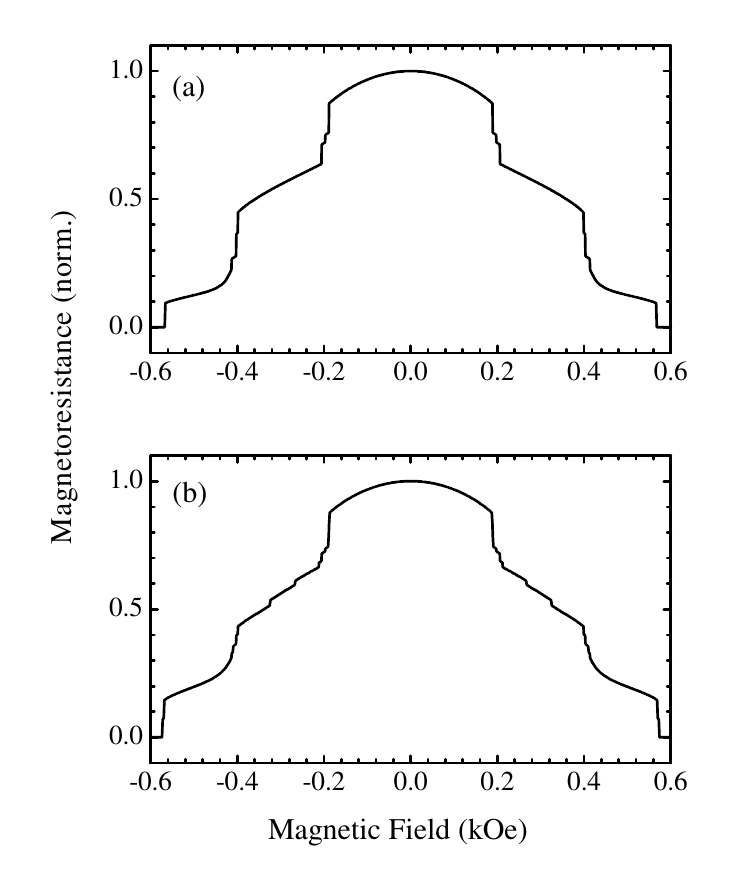}
\caption {Normalized magnetoresistance curves for growth direction [110] for the (a) fifth and (b) seventh Fibonacci generations.}
\label {fig3}
\end{figure}

\begin{figure}
\centering
\includegraphics[scale=1.5]{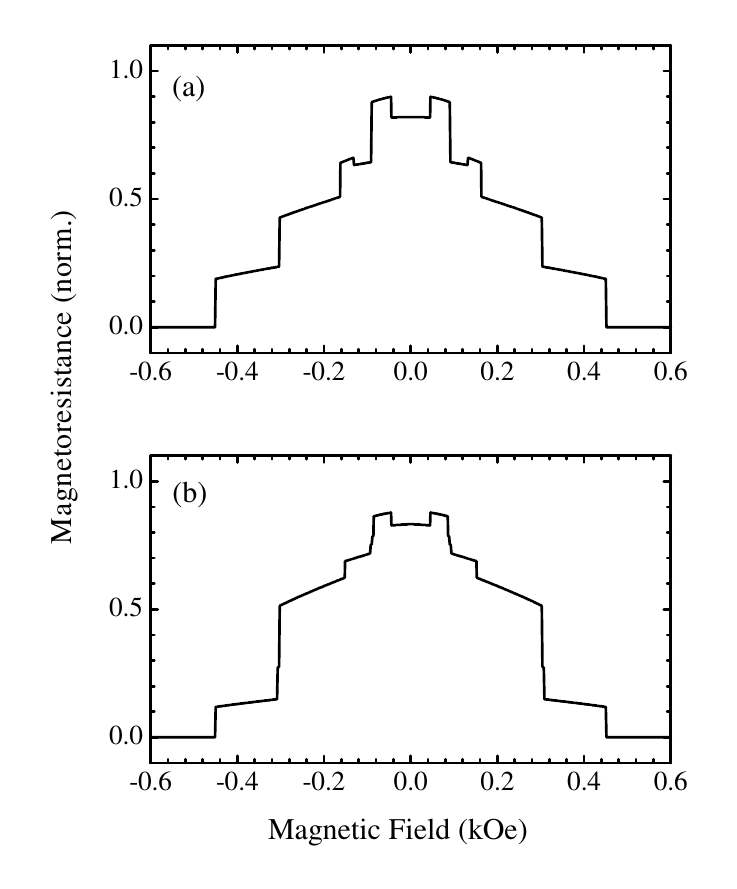}
\caption {Normalized magnetoresistance curves for the (a) fourth and (b) fifth Fibonacci generations obtained for the growth direction [100]. Positive magnetoresistance changes are evident.}
\label {fig4}
\end{figure}

\begin{figure}
\centering
\includegraphics[scale=0.6]{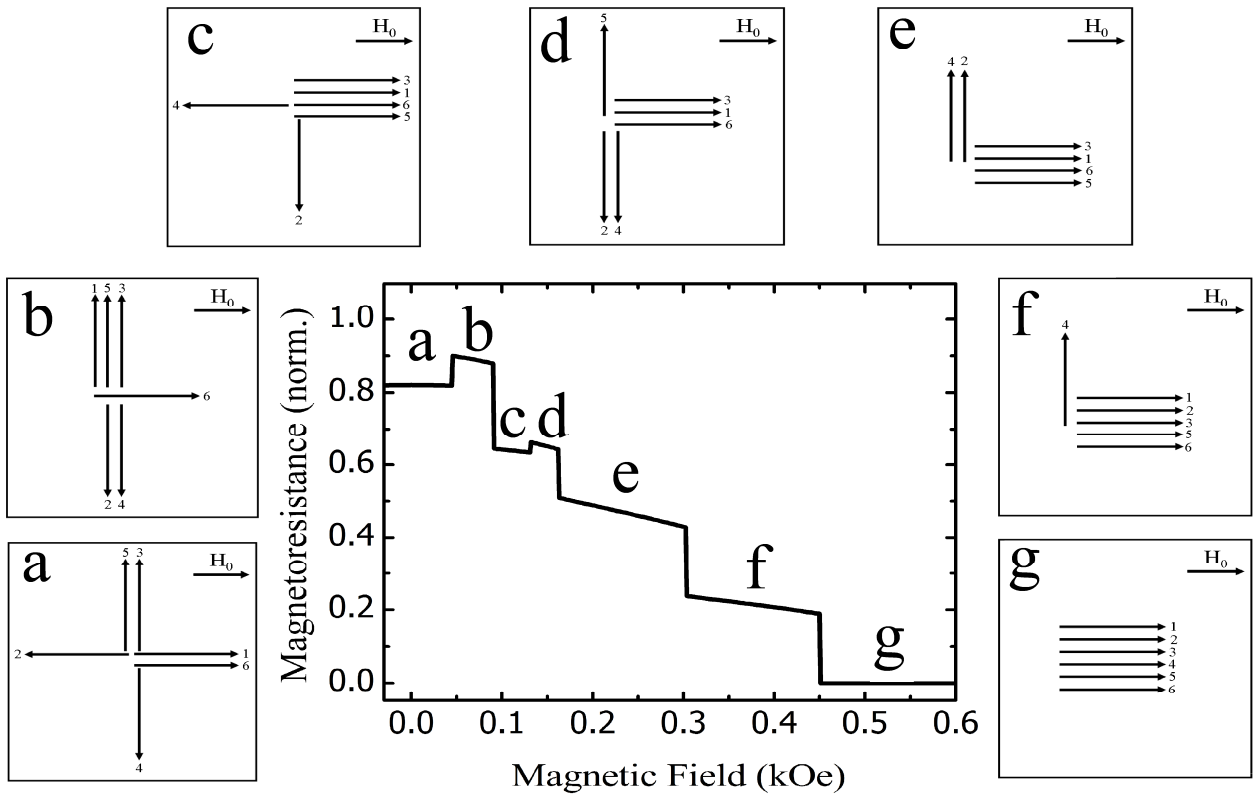}
\caption {Diagram of the fourth Fibonacci generation (growth direction [100]). The magnetization of each ferromagnetic layer is represented schematically by an arrow.}
\label {fig5}
\end{figure}

As illustrated in Figs.\ \ref{fig2} and \ref{fig3}, magnetoresistance curves for odd and even Fibonacci generations show different profiles. This is a consequence of the quasiperiodic sequence itself, since even generations of the sequence terminate with $B$, while odd generations start and end with $A$. This subtle difference is responsible for the well-defined even and odd parity related to the generation number of the Fibonacci structure. A similar parity had been observed previously in the specific heat of quasiperiodic magnetic super-lattices \cite{bezerra1}. On the other hand, a much more interesting and novel behavior is the positive change of magnetoresistance characterized by $\Delta M_R/\Delta H > 0$, illustrated in Figs.\ \ref{fig4} and \ref{fig5}. Our numerical results showed that in the low field region, the transitions, induced by the increase of the magnetic field, may lead to a magnetic configuration with a higher magnetoresistance. This is a direct consequence of the quasiperiodic distribution of the Cr layers in the multilayer structure.

We would like to thank the Brazilian Research Agencies CNPq, CAPES, FINEP and FAPERN for financial support. Work at Argonne was supported by the US Department of Energy, Basic Energy Sciences under contract No. DE-AC02-06CH1135.

{}

\newpage

\newpage

\newpage

\newpage

\newpage

\end{document}